\documentclass{aa}  

\usepackage{graphicx}
\usepackage{cleveref}
\usepackage{txfonts}
\usepackage{natbib}
\bibpunct{(}{)}{;}{a}{}{,} 
\newcommand{\Msun}{{M}_{\odot}}

\begin{document}

   \title{Accretion Flares from Stellar Collisions in Galactic Nuclei}


   \author{B. X. Hu
          \inst{1}
          \and
          A. Loeb\inst{2}
          }

   \institute{Department of Physics, Harvard University\\
              \email{bhu@g.harvard.edu}
         \and
             Department of Astronomy, Harvard University\\
             \email{aloeb@cfa.harvard.edu}
             }

   \date{Received Month Day, Year; accepted Month Day, Year}

 
  \abstract
   {The strong tidal force in a supermassive black hole's (SMBH) vicinity, coupled with a higher stellar density at the center of a galaxy, make it an ideal location to study the interaction between stars and black holes. Two stars moving near the SMBH could collide at a very high speed, which can result in a high energy flare. The resulting debris can then accrete onto the SMBH, which could be observed as a separate event.}
   {We simulate the light curves resulting from the fallback accretion in the aftermath of a stellar collision near a SMBH. We investigate how it varies with physical parameters of the system.}
   {Light curves are calculated by simulating post-collision ejecta as $N$ particles moving along individual orbits which are determined by each particle's angular momentum, and assuming that all particles start from the distance from the black hole at which the two stars collided. We calculate how long it takes for each particle to reach its distance of closest approach to the SMBH, and from there we add to it the viscous accretion timescale as described by the alpha-disk model for accretion disks. Given a timestamp for each particle to accrete, this can be translated into into a luminosity for a given radiative efficiency.}
   {With all other physical parameters of the system held constant, the direction of the relative velocity vector at time of impact plays a large role in determining the overall form of the light curve. One distinctive light curve we notice is characterized by a sustained increase in the luminosity some time after accretion has started. We compare this form to the light curves of some candidate tidal disruption events (TDEs).}
   {Stellar collision accretion flares can take on unique appearances that would allow them to be easily distinguished, as well as elucidate underlying physical parameters of the system. There exist several ways to distinguish these events from TDEs, including the much wider range of SMBH masses stellar collisions may exist around. The beginning of the Vera Rubin Observatory Legacy Survey of Space and Time (LSST) will greatly improve survey abilities and facilitate in the identification of more stellar collision events, particularly in higher-mass SMBH systems.}


   \maketitle
%

\section{Introduction}
\label{sec:1}

   Tidal disruption events (TDEs) were first theorized in the 1970s as the consequence of a star moving too close to a supermassive black hole (SMBH) and being torn apart by the SMBH's tidal forces \citep{hills1975, lidskii1979, carter1982, carter1983, stone2019, gezari2021}. The first TDEs were discovered in archival searches of the ROSAT All Sky Survey (RASS) conducted in 1990-1991 \citep{donley2002}, and in years since they have proven to be a powerful probe of SMBH physics, such as by indicating the presence of a quiescent SMBH through the subsequent energetic accretion flare \citep{rees1988}. To date, there exist several dozen events which are believed to be TDEs and the list is only expected to grow through increased survey power such as from the Vera Rubin Observatory Legacy Survey of Space and Time (LSST) \citep{ivezic2019} in the optical, and the extended Roengen Survey with an Imaging Telescope Array (eROSITA) \citep{predehl2021} in the X-ray. More recently, observations have revealed potential TDE events which display unusual types of behavior. 

   In this work we consider a separate event that can occur at the center of galaxies, and therefore which could be observed and incorrect interpreted as a TDE. A distinct phenomenon that can also occur due to stars moving close to the SMBH is a collision between two high-speed stars. This kind of collision has been studied in works such as \cite{rubin2011} and \cite{balberg2013}, which consider how the cluster of stars at the galactic center can build up and eventually reach a steady state condition in which the rate of collisions equals the formation rate of stars. In our previous work \cite{hu2021}, we considered how if this collision occurs sufficiently close to the SMBH, the resulting light curve can be very energetic but short-lived. We focused specifically on SMBH with $M_{\bullet} > 10^8\,M_{\odot}$, the reason being that for sun-like stars, the tidal disruption radius $R_T=(M_{\bullet}/M_*)^{1/3}R_*$ for a star with mass $M_*$ and radius $R_*$ is smaller than the black hole's event horizon radius $R_S=2GM_{\bullet}/c^2$ for black hole masses in this range. Therefore we can more easily study the highest velocity stars that would still result in visible phenomena, those which move closest to the event horizon radius. In our prior work we studied the light curve that would result from the collision itself and compared it to those from supernova explosions. In this work we examine how these collision events can likely result in a stream of debris that can accrete onto the SMBH and create an accretion flare that can resemble that from TDEs. Furthermore, we extend our previous analysis and consider smaller SMBHs in this work while still constraining these events to occur at radii beyond the TDE radius, not only because these smaller mass SMBHs are more common throughout the universe \citep{behroozi2019}, but also to draw comparisons between our theorized light curves and existing events that are believed to be TDEs. Like TDEs, stellar collisions collisions can fuel SMBH flares. For a collision rate of once per $10^5$ years \citep{rubin2011, hu2021}, there would be $\sim 10^5$ such events per SMBH during the age of the Universe. If the flares last a year, one in $\sim 10^5$ SMBHs will be lit by such a flare in any snapshot of the sky. This is roughly the same rate as that of TDEs.
   
   The outline of this paper is as follows. In section 2, we describe our method for arriving at the mass accretion rate and light curve from the stream of debris from a stellar collision. In section 3, we present examples of light curves while varying free parameters in our data. We also compare our results to light curves from six TDEs showing interesting features. In section 4, we discuss our results from running simulations. Finally, in section 5, we discuss conclusions and future work.

   \begin{figure*}
   \centering
   \includegraphics[width=0.8\textwidth]{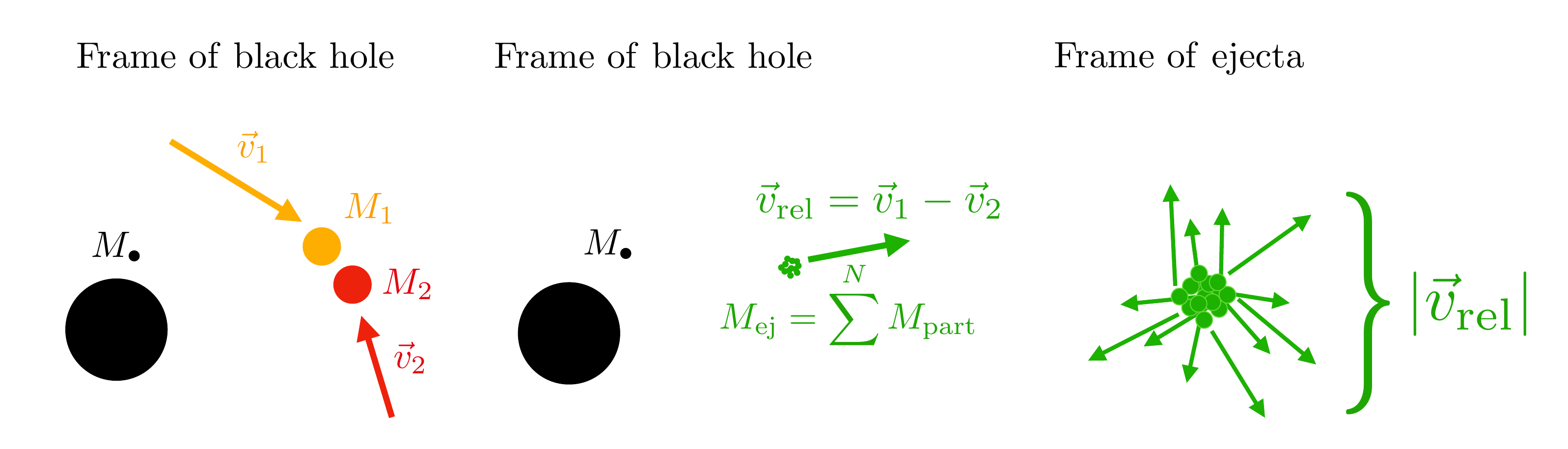}
   \caption{The geometry of the collision event. In panel (a) two stars of mass $M_1$ and $M_2$ approach each other with velocities $\vec{v}_1$ and $\vec{v}_2$ in the frame of the black hole, respectively. In panel (b) the ejecta has mass $M_{\mathrm{ej}}$ and consists of $N$ equal-mass particles. The ejecta moves around the black hole with relative velocity $\vec{v}_{\mathrm{rel}}$. In panel (c) we switch to the frame of the ejecta, in which each individual particle moves in a random direction with speed $|\vec{v}_{\mathrm{rel}}|$.}
              \label{fig:1}
    \end{figure*}

\section{Method}
\label{sec:2}

   We simulate the post-collision ejecta as $N$ particles of equal mass arranged in a spherical distribution. For numerical ease we consider ejecta with mass $1\,M_{\odot}$, and note that for smaller or larger ejecta the mass accretion rate and therefore luminosity simply scale linearly with the mass of the ejecta. \cite{benz1987} conduct fully three-dimensional calculations of collisions between identical stars and finds that in a grazing collision between two stars, at low enough speeds they are likely to form a spiral-shaped mass distribution which then proceeds to coalesce within a dynamical time. Mass loss then occurs through both direct ejection and formation of an accretion disk around the coalesced object. 

   We assume that the ejecta with mass $M_{\mathrm{ej}}$ continues moving around the SMBH with the same center-of-mass velocity as the two stars shortly before collision. This relative velocity is drawn from a Maxwellian distribution that describes a galaxy's stellar density profile $\rho_{\eta}(r_{\mathrm{gal}})$ provided by \cite{tremaine1994}, 
   \begin{equation}\label{eq:1}
    \ \rho_{\eta}(r_{\mathrm{gal}})\equiv\frac{\eta}{4\pi}\frac{r_sM_{\mathrm{\mathrm{sph}}}}{r_{\mathrm{gal}}^{3-\eta}(r_s+r_{\mathrm{gal}})^{1+\eta}},
    \end{equation}
    where we adopt the commonly used value $\eta=2$ \citep{hernquist1990}, $M_{\mathrm{sph}}$ is the total mass of the host spheroid, and $r_s$ is a distinctive scaling radius. We note that the relative velocity can vary between 0 and a fraction of the speed of light because we are not including special and general relativistic corrections. Although mathematically the relative velocity is not constrained at the upper end, we note that in a typical calculation with tens of thousands of samples, the maximum relative velocity observed is no more than $\sim5\%$ the speed of light. 

   An illustration of the configuration we consider is provided in Fig. \ref{fig:1}. The left panel shows the geometry immediately prior to collision; two stars with masses $M_1$, $M_2$ and velocities $\vec{v}_1$, $\vec{v}_2$, with respect to the black hole, collide. Their relative velocity with respect to the black hole is $\vec{v}_{\mathrm{rel}}=\vec{v}_1-\vec{v}_2$. Conservation of momentum dictates that ejecta will continue moving with this same relative velocity after the collision for $M_1=M_2$. For simplicity, we assume this is the case. The middle panel shows the geometry immediately post-collision. We assume our ejecta consists of $N$ particles of equal mass which total $M_{\mathrm{ej}}$. The ejecta continues moving at $\vec{v}_{\mathrm{ej}}$ with respect to the black hole. The right panel shows a close up of the ejecta from the middle panel, but now in the frame of the ejecta. Each of the $N$ particles that make up the ejecta move in a random direction with speed $|\vec{v}_{\mathrm{rel}}|$. Throughout our calculations, and as shown in Fig. \ref{fig:1}, for the purposes of obtaining order-of-magnitude estimates, we do not consider the effects of dynamical friction from other stars in the vicinity.
   
\begin{figure}
   \centering
   \includegraphics[width=8cm]{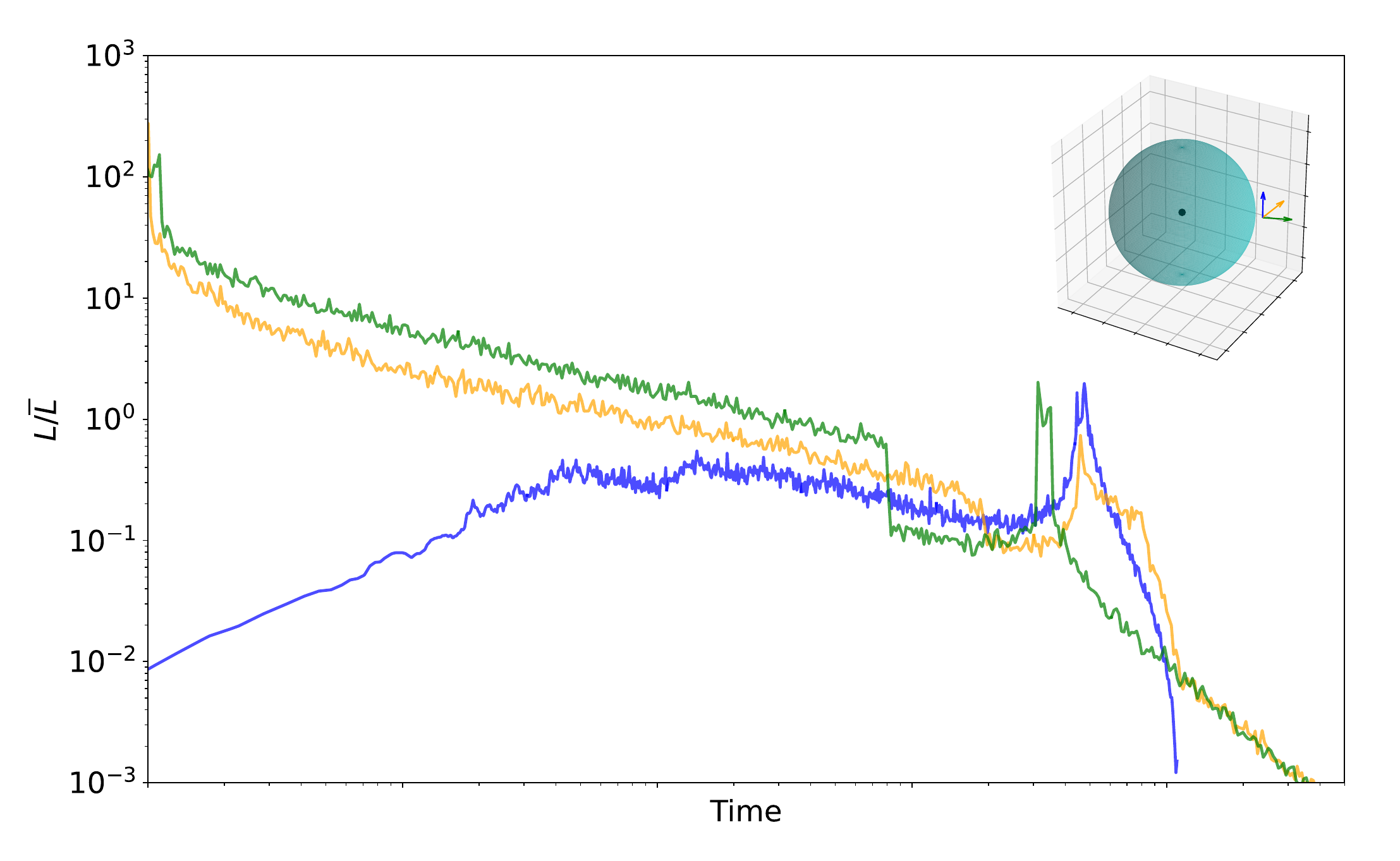}
      \caption{We simulate three different collision events with the relative velocity vectors pointed in three different directions, as shown in the upper right hand corner of the plot. On the $y$-axis we plot the luminosity $L$ divided by the average luminosity over the duration of the light curve $\bar{L}$, and on the x-axis we plot time. We intentionally exclude units on the x-axis because the timescales over which the accretion events occur can vary widely, as shown in Fig. \ref{fig:8}. We see that while the overall magnitude of the event is not significantly affected by the relative velocity vector orientation, the shape of the light curve, including peaks, rises, and falls, is affected.
              }
         \label{fig:2}
   \end{figure}

   In calculating this distribution it is assumed that the velocity vector has no preferred orientation. While this may be true, in our simulations the orientation of the velocity vector affects the overall form of the light curve. Therefore we must include the polar and azimuthal angles associated with the velocity vector as additional parameters. In the frame of the collision we also assume that the $N$ ejecta particles disperse in random directions, each with speed $v_{\mathrm{rel}}$. The end result is that from the frame of the viewer, for each of the $N$ particles we add together two velocity vectors of the same magnitude but with two different, randomly generated directions--one which describes the overall velocity of the ejecta and another which describes the velocity of the individual particle. From the viewer's perspective, particles travel at speeds ranging from 0 to $2v_{\mathrm{rel}}$ with an average value of $v_{\mathrm{rel}}$. In Fig. \ref{fig:2} we show an example where, with all other factors (mass of ejecta, mass of black hole, distance from black hole, magnitude of relative velocity vector) held constant, the angle of approach can have a significant effect on the overall shape of the light curve. In the corner of the plot, colored arrows indicate the relative velocity vector orientations with respect to the black hole, shown as a black dot at the center of the sphere.

%

   We assume that all particles start at the initial radius of the collision, $r_{\mathrm{init}}$. From there, we must determine what path they take towards the black hole and how long they take to accrete in order to calculate the mass accretion radius. If we consider the ejecta particles to be freely moving test particles under Schwarzschild geometry, then they move along geodesics of spacetime. In the simplest case of radial infall, the proper time $\tau$ for a particle to move from some radius starting $r$ to some final radius $R$ is,
    \begin{equation}\label{eq:2}
    \ \tau=\left(\frac{R^3}{8M_{\bullet}}\right)^{1/2}\left[2\left(\frac{r}{R}-\frac{r^2}{R^2}\right)^{1/2}+\cos^{-1}\left(\frac{2r}{R}-1\right)\right].
    \end{equation}
    We consider that some particles might initially move away from the black hole and then turn around. We assume that each particle starts out with some energy $E_{\mathrm{init}} = \frac{1}{2}v_{\mathrm{rel}}^2 - GM_{\bullet} / r_{\mathrm{init}}$. If this theoretical turn-around radius exists, it should occur when the particle comes to a stop, $r_{\mathrm{turn}} = -GM_{\bullet} / E_{\mathrm{init}}$. 

    In addition, we consider another important radius for the particle, the radius of closest approach. Given the particle's initial angular momentum $\vec{h}=\vec{r}\times\vec{v}$ this is described by,
    \begin{equation}\label{eq:3}
    \ r_{\mathrm{close}}=\frac{|\vec{h}|^2}{GM_{\bullet}\left(1+e\right)},
    \end{equation}
    with $e$ the eccentricity of the ejecta's path around the black hole.

    Finally, we must consider how the time for the particle to fall inwards is affected by the accretion disk. As a particle moves within the accretion disk, it moves in a turbulent flow, losing both energy and angular momentum. We adopt the conventional alpha-disk model \citep{shakura1973} using the $\alpha$ parameter, $\alpha\lesssim1$, which parametrizes the unknown effective increase in viscosity due to turbulent eddies within the accretion disk.  In the alpha-disk model, we parametrize the kinematic velocity due to turbulence as $\nu(R, t) = \alpha\,a(R, t)H(R, t)$, with $a(R, t)$ the sound speed and $H(R, t)$ the disk thickness. Observations can be used to constrain $\alpha$, which is typically calculated to vary between $\sim0.01$ and $\sim0.1$ \citep{lasota2023}. In our work, we adopt $\alpha=0.1$. At some radius $R$, we estimate the viscous accretion timescale as $t_{\mathrm{acc}} = R^2 / \nu = R^2 / \alpha\,a\,H$. We estimate $H\sim a/\Omega_k$ in an alpha disk, with $\Omega_k=\left(\vec{r}\times\vec{v}\right)/r^2$ the angular velocity. Finally, we estimate the sound speed as $a\sim v/2$, with $v$ the Keplerian velocity, arriving at,
    \begin{equation}\label{eq:4}
    \ t_{\mathrm{acc}} = \frac{4R^2\left|\vec{r}\times\vec{v}\right|}{\alpha\,r^2\,v^2},
    \end{equation}
    with $R$ the radius of closest approach and $\vec{r}$ and $\vec{v}$ the initial radius and velocity vectors of the particle. In our approach, we ignore the effect of gas pressure along the path of the debris towards the SMBH until the ejecta impacts the accretion disk. This is an approximate model that can be tested by hydrodynamic simulations in future work.

     The final procedure for calculating the particle's time to accrete is as follows: we first check if the particle initially moves away from the SMBH before coming to a stop and turning around. If it does, we calculate the time this first segment takes assuming the particle moves along a geodesic of spacetime, i.e. that the effects of the accretion disk may be negligible in the immediate aftermath of the collision. Following this first step, all particles in the calculation are either at this turn-around radius or, if they don't turn around, then they are still at the radius of the collision. We then calculate the time it takes for the particle to move to the radius of closest approach, again along a geodesic of spacetime. Finally, at the radius of closest approach, we calculate the corresponding viscous accretion timescale as given by the above equation. We add all the timescales together for each particle to arrive at its time to accrete. 

     Knowing each of the $N$ particle's mass and accretion timescale, we can translate this to a mass accretion rate. To translate the mass accretion rate to the luminosity, we need the radiative efficiency. Because we are interested in comparing our final light curves to various TDEs with light curves in both the radio and optical regimes, we first estimate the fraction of energy we expect to be emitted in the radio frequency range. \cite{cendes2022} studied TDE AT2018hyz flux density versus frequency measurements from VLA+ALMA, Chandra, UVOT, and ZTF data. This was integrated to find that approximately $5\%$ of the energy is emitted in radio. For optical light curves, we adopt a value of $10\%$ \citep{duras2020}. We then consider the radiative efficiency of hot accretion flows. As described by \cite{xie2012}, we know that only a small fraction of the matter in the accretion disk will end up falling onto the black hole, and that electrons can receive a large fraction of the viscously dissipated energy in the accretion flow. \cite{xie2012} provides a systematic calculation that takes in account both these considerations and provides fitting formulae of radiative efficiency as a function of accretion rate for various $\delta$, the fraction of turbulent dissipation that heats the electrons directly,
    \begin{equation}\label{eq:xie}
    \ \epsilon(\dot{M}_{\mathrm{net}})=\epsilon_0\left(\frac{\dot{M}_{\mathrm{net}}}{\dot{M}_{\mathrm{c}}}\right)^a,
    \end{equation}
    with $\dot{M}_{\mathrm{net}}$ the accretion rate at the Schwarzschild radius, $\dot{M}_{\mathrm{c}}=10^{-2}\dot{M}_{\mathrm{Edd}}=10^{-1}L_{\mathrm{Edd}}/c^2$, and $\epsilon_0,a$ fitting constants provided in Table 1 of \cite{xie2012}.
    
     Knowing each particle's mass, time to accrete, the fraction of energy emitted in radio, and the radiative efficiency, we can calculate the luminosity associated with the fallback of the ejecta mass after the collision. However, we note that there is one more free parameter, which is the mass of the ejecta. At the start of calculations it is taken as $1\,M_{\odot}$ for simplicity, but we note that the mass accretion rate scales linearly as the mass of the ejecta changes. While the luminosity would not necessarily scale linearly because the radiative efficiency is better described as multiple piecewise power-law equations that depend on the mass accretion rate, as long as the mass of the ejecta does not change by more than an order of magnitude, the change in the radiative efficiency rate is likely to have little effect, and we can generally guess that the new light curve can also be approximated by scaling the original light curve linearly with the change in the ejecta mass.


\section{TDE-like Flares of Interest}
\label{sec:3}

    In this work we consider multiple potential TDEs studied in recent years, some of which have been noted for their unusual light curves. TDEs occur when the tidal force from the SMBH at some distance $r$ from the SMBH for a star of mass $M_*$ and radius $R_*$, $GM_{\bullet} R_* / r^3$, overwhelms the self-gravity of the star, $GM_*/R_*^2$, which roughly occurs at some distance $R_T=(M_{\bullet}/M_*)^{1/3}R_*$ known as the TDE radius \citep{hills1975, rees1988}. Furthermore, TDEs are expected to only be observable when the event happens outside of the SMBH's Schwarzschild radius, which constrains observable events to occur around black holes with mass $M_{\bullet} < 10^8\,M_{\odot}$ for nonspinning black holes \citep{stone2019}. The fallback timescale, defined as the orbital period of the most bound debris, follows $t_{fb} = 2\pi GM_{\bullet}(2E)^{-3/2}$, with $E$ the energy of the most bound debris. If we further assume that the specific energy distribution is uniform, $dE/dM = 0$, we can derive that the fallback rate of TDE debris follows a power law, $dM/dt\propto(t-t_D)^{-5/3}$ \citep{gezari2021}. Remarkably, this simply derived power law has been well-observed in optical light curves.  Radio emission from a TDE was first detected during a multiwavelength campaign for the low-redshift optically selected TDE ASASSN-14li \citep{jose2014, vanvelzen2016}. There are many interpretations of the radio emission, including synchrotron emission from external shocks \citep{alexander2016}, the unbound debris stream \citep{krolik2016}, a relativistic jet \citep{vanvelzen2016b}, and internal shocks in a relativistic jet \citep{pasham2018}. However, the detection of a correlation between the X-ray and radio light curves, $L_{\mathrm{radio}}\propto L_X^2\propto M^dot_acc^2$ \citep{pasham2018} suggests that accretion and jet power are coupled, favoring the aforementioned internal jet model. Radio follow-up has shown that high radio-luminosity, relativistic jets occur in around $\sim1\%$ of TDEs \citep{alexander2020}.

\begin{figure}
   \centering
   \includegraphics[width=8cm]{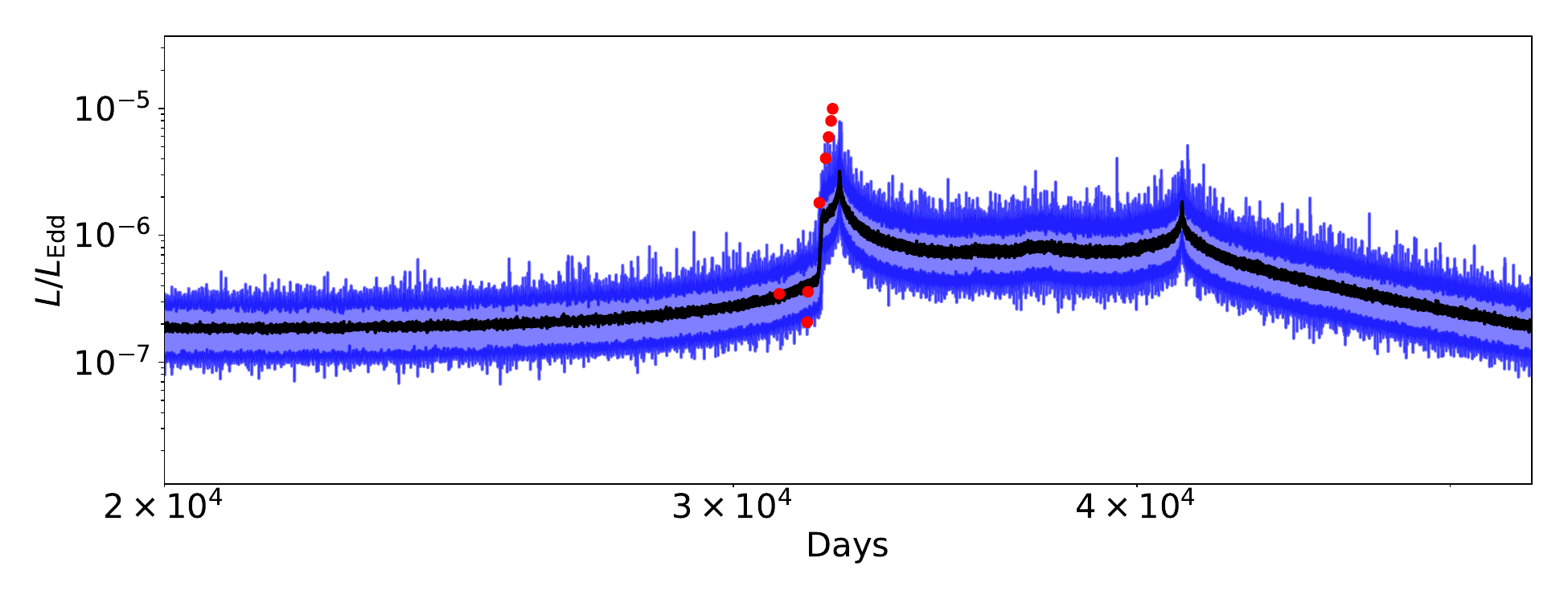}
      \caption{The best-fitting simulated light curve to the AT2018hyz data is shown in black. Overlaid in blue are the $\pm1,2\sigma$ confidence bands.}
         \label{fig:3}
   \end{figure}

    TDE AT2018hyz was first detected by the All-Sky Automated Survey for Supernovae (ASAS-SN) \citep{shappee2014, kochanek2017} on October 14, 2018. The UV-Optical Telescope (UVOT) \citep{roming2005} on board the Neil Gehrels Swift observatory \citep{gehrels2004} subsequently observed AT2018hyz from November 10, 2018 until July 8, 2019. Four hours of AMI-LA observations on November 15, 2018, 32 days after optical discovery, revealed no radio source at the location of the transient. The upper limit of the observation corresponds to a luminosity of $L_{\nu} < 6.6\times10^{37}\,\mathrm{erg}/\mathrm{s}$. The transient is located in the nucleus of the galaxy 2MASS J10065085+0141342 located at redshift $z = 0.04573$. Relative astrometry has been performed that the transient is nuclear, with a separation of $0.2\pm0.8\,\mathrm{kpc}$. AT2018hyz was then observed at 972 days with the Karl G. Jansky Very Large Array (VLA) as part of a study of late-time radio emission from TDEs, which succeeded in detecting a source. Multi-frequency observations were then conducted from L- to K-band. This observed radio emission data, along with some collected by the VLA Low-band Ionosphere and Transient Experiment (VLITE), the MeerKAT radio telescope, and the Australian Telescope Compact Array (ATCA), is shown in the light curve in Fig \ref{fig:3}. As can be seen in the figure, the radio light curve shows the unusual feature of increasing rapidly from $7\times10^{37}\,\mathrm{erg}/\mathrm{s}$ to $2\times10^{39}\,\mathrm{erg}/\mathrm{s}$ over $\sim600$ days. \cite{gomez2020} uses the TDE package in the Python package MOSFiT to model the light curves and estimate a black hole mass of $5.2\times10^6\,M_{\odot}$, which is in agreement with similar results found in Hung et al. 2020 \citep{hung2020}. In addition, the stellar velocity dispersion measured from the SDSS spectrum gives, using the $M_{\bullet}-\sigma_*$ relation from \citep{xiao2011}, can be used to estimate a black hole mass of $1.6\times10^6\,M_{\odot}$. Given these various roughly consistent results, we adopt the last mass value in our work. 

   The second TDE-like flare we consider in this work is ASASSN-15oi. Observations of this transient from ASAS-SN first triggered the detection pipeline on August 14, 2015 \citep{brimacombe2015}, from a location within the galaxy 2MASX J20390918-3045201. UVOT conducted follow-up observations in the UV and X-ray Telescope (XRT) \citep{burrows2005} confirmed its detection in the X-rays. \cite{holoien2016} fits stellar population synthesis models to archival GALEX data, ASAS-SN V-band measurements, and 2MASS JHK$\_$S magnitudes of the host galaxy to find a stellar mass of $M_*=1.1\times10^{10}\,M_{\odot}$, which implies a bulge mass of $6.3\times10^9\,M_{\odot}$ based on the average stellar-mass-to-bulge-mass ratio from ASASSN-14ae and ASASSN-14li, which in turns implies a black hole mass of $1.3\times10^7\,M_{\odot}$ using the $M_{\mathrm{bulge}}-M_{\bullet}$ relation \citep{mcconnell2013}. More recently, \cite{gezari2017} and \cite{mockler2019} arrived at a lower mass estimate of $2.5\times10^6\,M_{\odot}$ using the aforementioned package MOSFiT. Immediately following optical discovery, VLA was used to carry out radio observations on ASASSN-15oi. Initial observations resulted in null detections. However, observations continued, motivated by theoretical models that suggest delayed signals may occur depending on how long the formation of the accretion disk takes. Null detections were recorded again at 23 and at 90 days. Significant radio emission was first detected at 182 days on February 12, 2016. Subsequently, a follow-up observation campaign was carried out in multiple radio frequencies, the results of which are displayed in figure AA. While the delayed radio flare at 182 days is surprising in itself, a second, even more surprising radio flare has since been detected years after the initial discovery in the optical. Various models have been considered to explain this unusual delayed emission, such as a standard CNM shockwave. model, a relativistic jet, and off-axis jet, and more. The existence of the second, even brighter flare makes all of these scenarios quite unlikely. It has been suggested that the rebrightening could be driven by whatever source caused the initially delayed radio emission, repeated partial disruptions of a star, or highly variable accretion due to the presence of a potential binary SMBH system. These proposed explanations still need to be tested against the available data.

    \begin{figure}
   \centering
   \includegraphics[width=8cm]{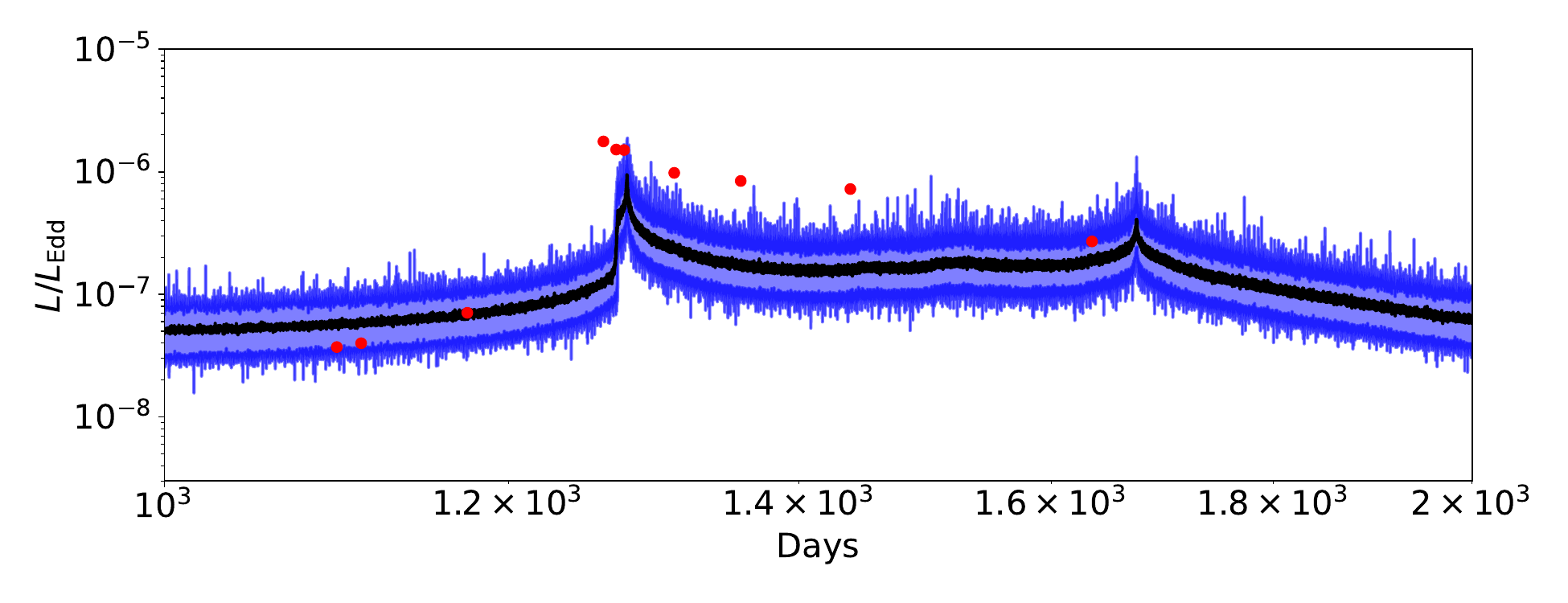}
      \caption{The best-fitting simulated light curve to the ASASSN-15oi data is shown in black. Overlaid in blue are the $\pm1,2\sigma$ confidence bands.}
         \label{fig:4}
   \end{figure}

\begin{figure}
   \centering
   \includegraphics[width=8cm]{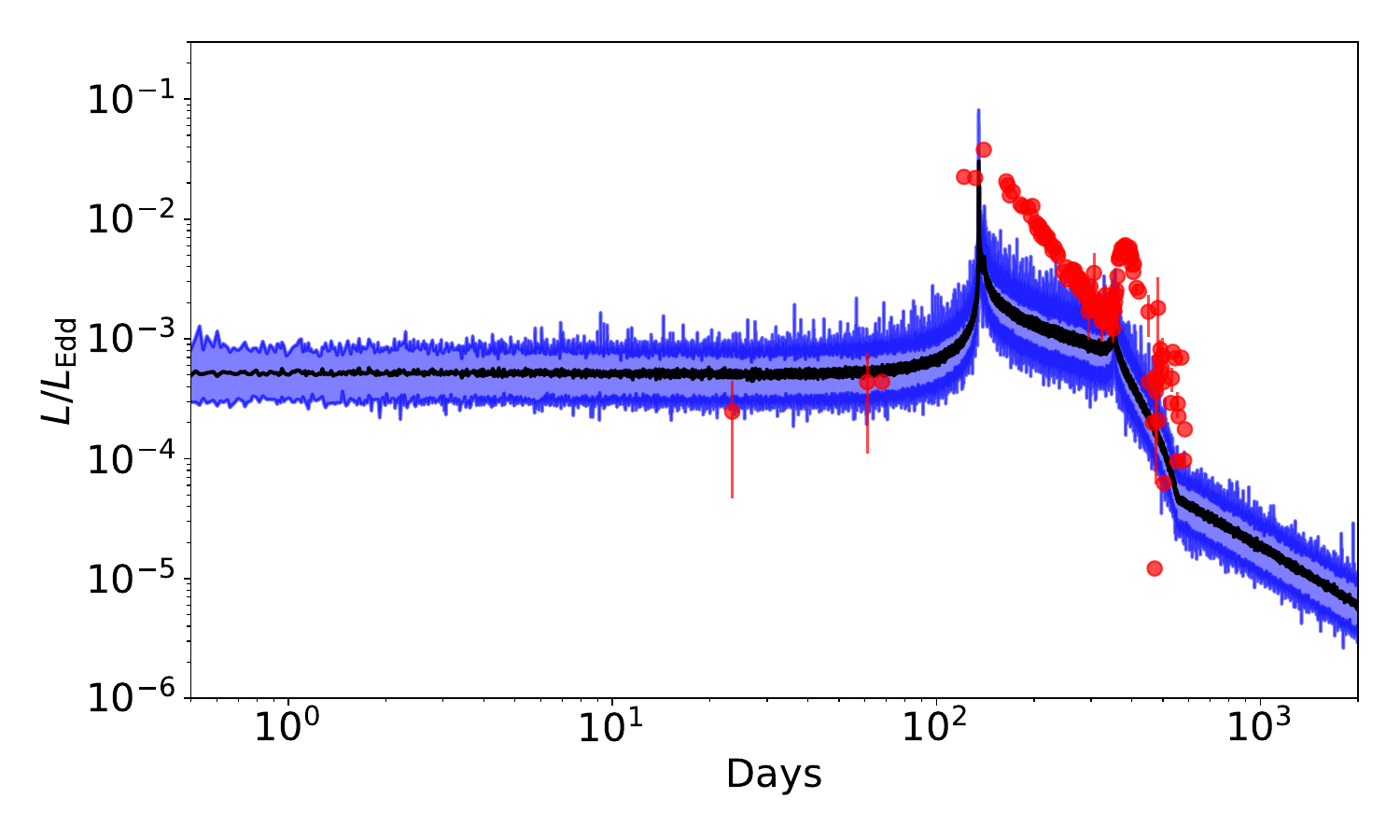}
      \caption{The best-fitting simulated light curve to the AT2019baf data is shown in is shown in black. Overlaid in blue are the $\pm1,2\sigma$ confidence bands.}
         \label{fig:5}
   \end{figure}

\begin{figure}
   \centering
   \includegraphics[width=8cm]{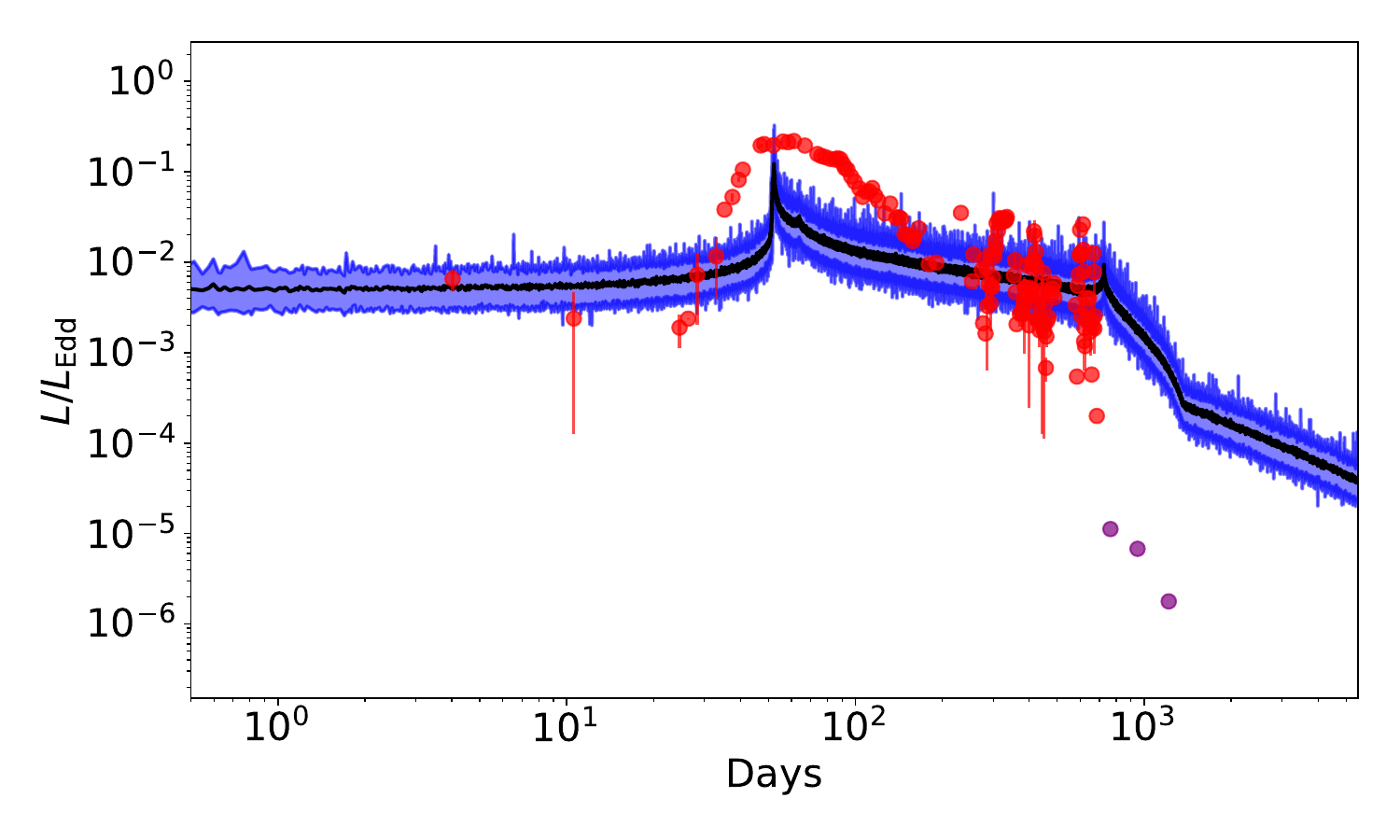}
      \caption{The best-fitting simulated light curve to the AT2019ehz data is shown in is shown in black. Overlaid in blue are the $\pm1,2\sigma$ confidence bands. In addition, late radio emissions are shown in purple.}
         \label{fig:6}
   \end{figure}

\begin{figure}
   \centering
   \includegraphics[width=8cm]{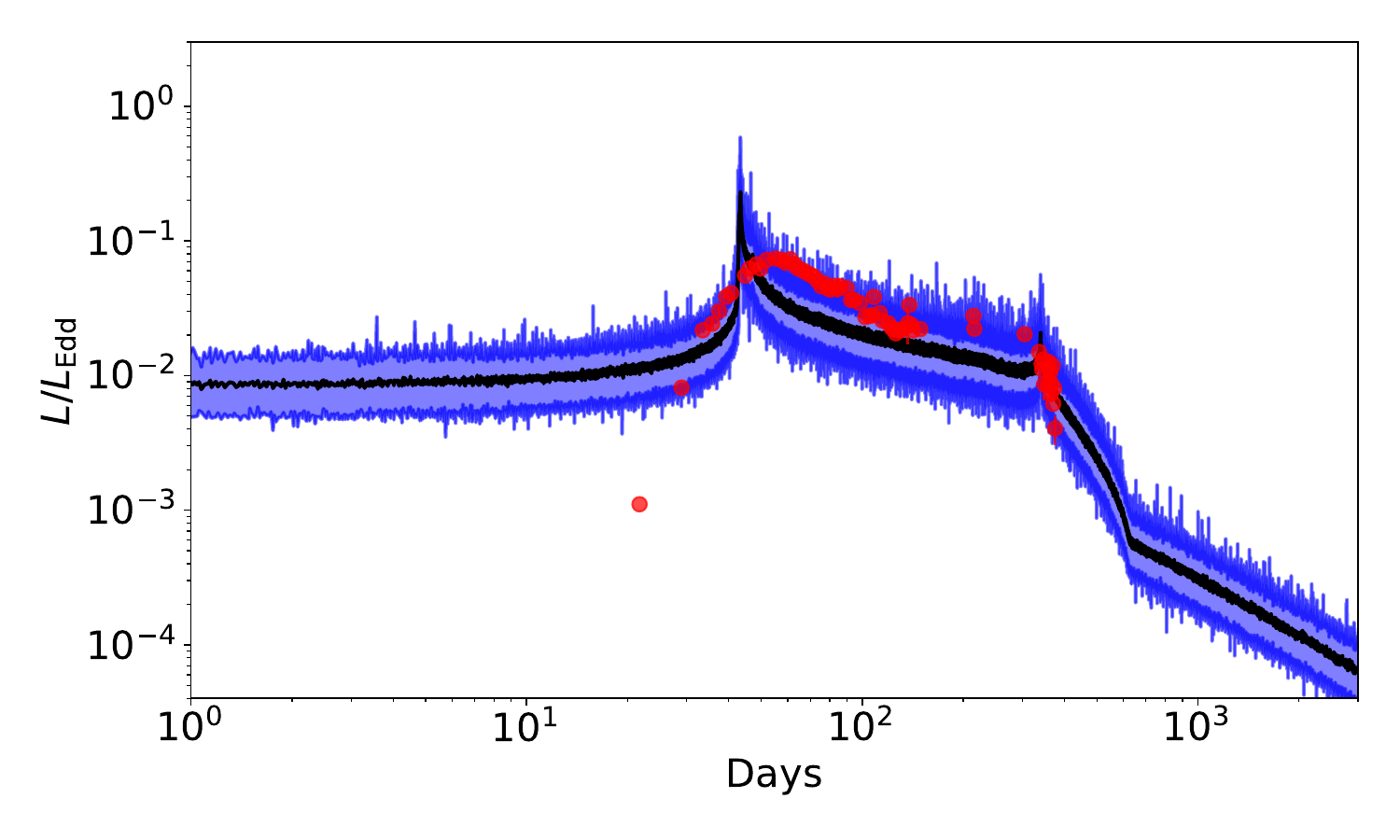}
      \caption{The best-fitting simulated light curve to the AT2019ehz data is shown in is shown in black. Overlaid in blue are the $\pm1,2\sigma$ confidence bands.}
         \label{fig:7}
   \end{figure}

   In addition, we consider three UV and optical light curves composed of data gathered from ZTF, the Ateroid Terrestrial-impact Last Alert System (ATLAS) \citep{tonry2018}, and the Ultra-Violet/Optical Telescope (UVOT) \citep{roming2005} on board the Neil Gehrels Swift Observatory \citep{gehrels2004}. The events we consider are AT2019baf, AT2019ehz, and AT2021uqv. For AT2019ehz, additional late radio emission is provided by \cite{cendes2023}. From table 5 in \cite{yao2023}, we adopt black hole masses of $\log (M_{\bullet}/\Msun)=6.89,\,5.81,\,$ and $6.27$, respectively, which are inferred either through $M_{\bullet}-\sigma_*$ relations or through $M_{\bullet}-M_{\mathrm{gal}}$ relations if $\sigma_*$ measurements are not available. Their light curves, plotted as dark blue dots, are overlaid with our best-fit models, plotted as light blue curves, in Figs \labelcref{fig:5,fig:6,fig:7}

\section{Analysis of Results}
\label{sec:4}

    We find that the accretion of the post-collision ejecta onto the black hole results in a light curve that can take on a varied appearance depending on several factors including the mass of the black hole, distance from the black hole at the time of collision, and direction of relative velocity vector. In this work, we focus primarily on one particular light curve form which is characterized by a an initial decay in the luminosity followed by a rapid rise, another decay in the luminosity, one final smaller rise in the luminosity, and the final decay. The most distinctive feature of the complete light curve is the sustained rise in luminosity between the starting and ending periods of decaying luminosity. We refer to the duration of this sustained rise in luminosity as the "peak-to-peak timescale" $t_{pp}$.

    We are able to track the orbital parameters of particles as they accrete onto the black hole. Broadly speaking, we find that the first wave of particles accreting onto the black hole are those with orbits oriented directly towards the black hole. These are followed by particles on hyperbolic orbits ($e>1$). Finally, particles that initially moved away from the black hole but eventually moved back towards the black hole are accreted. 
    
\begin{figure*}
   \centering
    \includegraphics[width=0.8\textwidth]{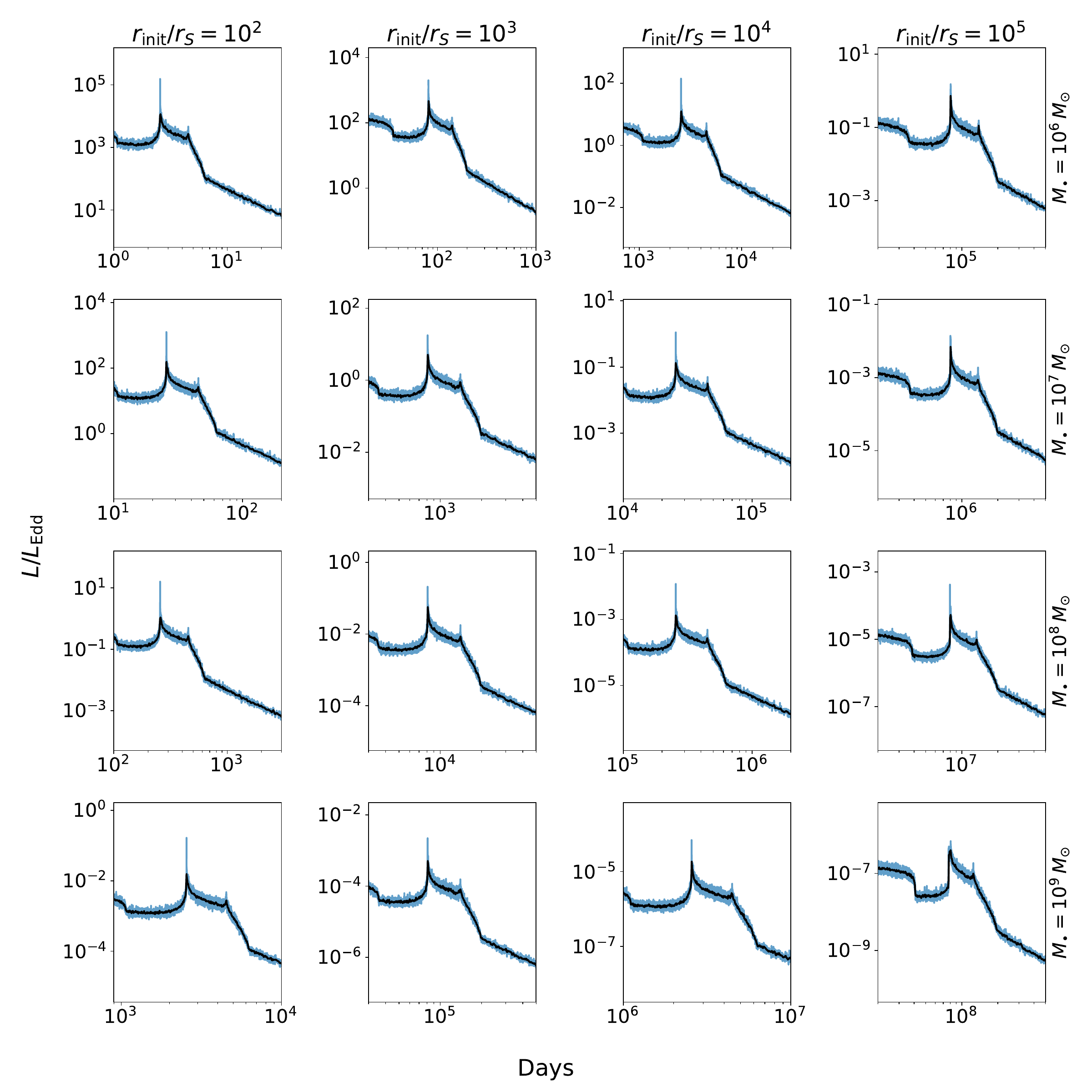} 
    \caption{Example light curves from stellar collision debris accreting onto a BH. Along the horizontal axis, we vary the distance of the collision from the black hole, normalized by the Schwarzschild radius, $r_{\mathrm{init}}/r_S$. Along the vertical axis, we vary the mass of the supermassive black hole, $M_{\bullet}$. The $y-$axis of every plot is normalized by the Eddington luminosity for the black hole, $L_{\mathrm{Edd}}$.} For all light curves shown here, the orientation is that shown by the yellow relative velocity vector in Fig. \ref{fig:2}.
    \label{fig:8}
    \end{figure*}

    Furthermore, we find that the duration of the increased luminosity period, $t_{pp}$, scales with two other physical parameters of the system: 1) the distance the ejecta starts from the black hole divided by the Schwarzschild radius, $r_{\mathrm{init}}/r_S$, and 2) the mass of the SMBH, $M_{\bullet}$. When one of the two parameters is held constant and the other is allowed to vary, there exists a power law relation between $t_{pp}$ and the other parameter. We find that $\log(t_{pp})\propto\log(r_{\mathrm{init}}/r_S)$ and $\log(t_{pp})\propto\log(M_{\bullet})$. These two relationships likewise follow the scaling relations that can be derived from the free-fall timecale as derived from Kepler's third law, $t_{\mathrm{orbit}}=\pi R^{3/2}/\sqrt{2G(M+m)}$, after substituting in the Schwarzschild radius for $R$ to derive the correct relation with respect to $M_{\bullet}$. Examples of light curves with the this distinctive structure are shown in Fig \ref{fig:8}. Along the x-axis, we vary $r_{\mathrm{init}}/r_S$. Along the y-axis, we vary $M_{\bullet}$. The angle of the relativity velocity vector is kept constant in all subplots. As can be seen in the figure, the duration of the flare, $t_{pp}$, increases with both parameters. Given the overall reliance of $t_{pp}$ on $M_{\bullet}$, $r_{\mathrm{init}}/r_S$, and the direction of the relative velocity vector, knowing two of the three parameters (most likely the first two) could for determination of the third.

\section{Conclusions \& Future Work}
\label{sec:5}
    In this work we simulate light curves from stellar collision debris accreting onto a nearby SMBH, characterize the light curves, and compare them to observed TDE light curves. We consider TDEs in this work because, like the high-speed stellar collisions we are interested in, they occur at the centers of galaxies, making it possible for the two events to be confused and misidentified. 
    
    We identify that although the light curves may have a wide variety of appearances, some have the distinctive trait of a sudden sustained rise in luminosity, the duration of which can be predicted based on the mass of the SMBH and the distance from the black hole the ejecta started at. We show examples of how our simulated light curves can resemble observed data that has previously been classified as TDE events. In future work, we plan on more systematically fitting our model to TDE light curves from existing datasets to identify potential stellar collision candidates.

    Although TDEs and stellar collisions can exist under similar circumstances, we highlight some important ways in which they can be distinguished. The stellar collision accretion flares described in this work are expected to be immediately preceded by a shorter, luminous flare as described in \cite{hu2021}. Partial tidal disruption events can result in repeated flares that can resemble the sustained flares in the light curves described in this work. However, partial tidal disruption events can also occasionally result in more than two flares from the continued disruption of the TDE debris \citep{bao2023, miniutti2023}. This long-term repeated behavior has not yet been demonstrated in stellar collision events. 
    
    Finally, TDEs are not expected to be observed in galaxies with SMBHs with $M_{\bullet}\gtrsim10^8 M_{\odot}$ for nonspinning black holes \citep{stone2019}, or $M_{\bullet}\gtrsim7\times10^8 M_{\odot}$ for maximally spinning black holes \citep{kesden2012}, due to the TDE radius being too low. Stellar collisions can continue to occur in galaxies with SMBH masses this high, which means that an observation of the type described in this work in a galaxy with such a high-mass SMBH is unlikely to have originated from a TDE. Although higher mass SMBHs are less populous in our Universe \citep{behroozi2019}, we will soon experience a substantial enhancement in survey power with the beginning of the Vera Rubin Observatory Legacy Survey of Space and Time (LSST) \citep{ivezic2019}.

\begin{acknowledgements}
      We thank Hamsa Padmanabhan for providing constructive comments and discussion. B.X.H. and A.L. acknowledge support from the Black Hole Initiative, which is supported by the John Templeton Foundation and the Gordon and Betty Moore Foundation. B.X.H. acknowledges support from the Department of Defense National Defense Science and Engineering Graduate Fellowship. 
\end{acknowledgements}

%
%

\bibliographystyle{aa}
\bibliography{ref}{}

\begin{thebibliography}{48}
\expandafter\ifx\csname natexlab\endcsname\relax\def\natexlab#1{#1}\fi

\bibitem[{{Alexander} {et~al.}(2016){Alexander}, {Berger}, {Guillochon},
  {Zauderer}, \& {Williams}}]{alexander2016}
{Alexander}, K.~D., {Berger}, E., {Guillochon}, J., {Zauderer}, B.~A., \&
  {Williams}, P.~K.~G. 2016, \apjl, 819, L25

\bibitem[{Alexander {et~al.}(2020)Alexander, van Velzen, Horesh, \&
  Zauderer}]{alexander2020}
Alexander, K.~D., van Velzen, S., Horesh, A., \& Zauderer, B.~A. 2020, Space
  Science Reviews, 216, 81

\bibitem[{Balberg {et~al.}(2013)Balberg, Sari, \& Loeb}]{balberg2013}
Balberg, S., Sari, R., \& Loeb, A. 2013, Monthly Notices of the Royal
  Astronomical Society: Letters, 434, L26

\bibitem[{{Bao} {et~al.}(2023){Bao}, {Guo}, {Zhang}, {Cheng}, {Yao}, {Li},
  {Yuan}, {Wang}, {Tsai}, \& {Chen}}]{bao2023}
{Bao}, D.-W., {Guo}, W.-J., {Zhang}, Z.-X., {et~al.} 2023, arXiv e-prints,
  arXiv:2311.16726

\bibitem[{{Behroozi} {et~al.}(2019){Behroozi}, {Wechsler}, {Hearin}, \&
  {Conroy}}]{behroozi2019}
{Behroozi}, P., {Wechsler}, R.~H., {Hearin}, A.~P., \& {Conroy}, C. 2019,
  \mnras, 488, 3143

\bibitem[{{Benz} \& {Hills}(1987)}]{benz1987}
{Benz}, W. \& {Hills}, J.~G. 1987, \apj, 323, 614

\bibitem[{{Brimacombe} {et~al.}(2015){Brimacombe}, {Brown}, {Holoien},
  {Stanek}, {Kochanek}, {Simonian}, {Basu}, {Beacom}, {Thompson}, {Shappee},
  {Prieto}, {Bersier}, {Dong}, {Falco}, {Wozniak}, {Szczygiel}, {Pojmanski},
  {Kiyota}, \& {Masi}}]{brimacombe2015}
{Brimacombe}, J., {Brown}, J.~S., {Holoien}, T.~W.~S., {et~al.} 2015, The
  Astronomer's Telegram, 7910, 1

\bibitem[{{Burrows} {et~al.}(2005){Burrows}, {Hill}, {Nousek}, {Kennea},
  {Wells}, {Osborne}, {Abbey}, {Beardmore}, {Mukerjee}, {Short}, {Chincarini},
  {Campana}, {Citterio}, {Moretti}, {Pagani}, {Tagliaferri}, {Giommi},
  {Capalbi}, {Tamburelli}, {Angelini}, {Cusumano}, {Br{\"a}uninger}, {Burkert},
  \& {Hartner}}]{burrows2005}
{Burrows}, D.~N., {Hill}, J.~E., {Nousek}, J.~A., {et~al.} 2005, \ssr, 120, 165

\bibitem[{{Carter} \& {Luminet}(1982)}]{carter1982}
{Carter}, B. \& {Luminet}, J.~P. 1982, \nat, 296, 211

\bibitem[{{Carter} \& {Luminet}(1983)}]{carter1983}
{Carter}, B. \& {Luminet}, J.~P. 1983, \aap, 121, 97

\bibitem[{{Cendes} {et~al.}(2023){Cendes}, {Berger}, {Alexander}, {Chornock},
  {Margutti}, {Metzger}, {Wieringa}, {Bietenholz}, {Hajela}, {Laskar}, {Stroh},
  \& {Terreran}}]{cendes2023}
{Cendes}, Y., {Berger}, E., {Alexander}, K.~D., {et~al.} 2023, arXiv e-prints,
  arXiv:2308.13595

\bibitem[{{Cendes} {et~al.}(2022){Cendes}, {Berger}, {Alexander}, {Gomez},
  {Hajela}, {Chornock}, {Laskar}, {Margutti}, {Metzger}, {Bietenholz},
  {Brethauer}, \& {Wieringa}}]{cendes2022}
{Cendes}, Y., {Berger}, E., {Alexander}, K.~D., {et~al.} 2022, \apj, 938, 28

\bibitem[{{Donley} {et~al.}(2002){Donley}, {Brandt}, {Eracleous}, \&
  {Boller}}]{donley2002}
{Donley}, J.~L., {Brandt}, W.~N., {Eracleous}, M., \& {Boller}, T. 2002, \aj,
  124, 1308

\bibitem[{{Duras, F.} {et~al.}(2020){Duras, F.}, {Bongiorno, A.}, {Ricci, F.},
  {Piconcelli, E.}, {Shankar, F.}, {Lusso, E.}, {Bianchi, S.}, {Fiore, F.},
  {Maiolino, R.}, {Marconi, A.}, {Onori, F.}, {Sani, E.}, {Schneider, R.},
  {Vignali, C.}, \& {La Franca, F.}}]{duras2020}
{Duras, F.}, {Bongiorno, A.}, {Ricci, F.}, {et~al.} 2020, A\&A, 636, A73

\bibitem[{{Gehrels} {et~al.}(2004){Gehrels}, {Chincarini}, {Giommi}, {Mason},
  {Nousek}, {Wells}, {White}, {Barthelmy}, {Burrows}, {Cominsky}, {Hurley},
  {Marshall}, {M{\'e}sz{\'a}ros}, {Roming}, {Angelini}, {Barbier}, {Belloni},
  {Campana}, {Caraveo}, {Chester}, {Citterio}, {Cline}, {Cropper}, {Cummings},
  {Dean}, {Feigelson}, {Fenimore}, {Frail}, {Fruchter}, {Garmire}, {Gendreau},
  {Ghisellini}, {Greiner}, {Hill}, {Hunsberger}, {Krimm}, {Kulkarni}, {Kumar},
  {Lebrun}, {Lloyd-Ronning}, {Markwardt}, {Mattson}, {Mushotzky}, {Norris},
  {Osborne}, {Paczynski}, {Palmer}, {Park}, {Parsons}, {Paul}, {Rees},
  {Reynolds}, {Rhoads}, {Sasseen}, {Schaefer}, {Short}, {Smale}, {Smith},
  {Stella}, {Tagliaferri}, {Takahashi}, {Tashiro}, {Townsley}, {Tueller},
  {Turner}, {Vietri}, {Voges}, {Ward}, {Willingale}, {Zerbi}, \&
  {Zhang}}]{gehrels2004}
{Gehrels}, N., {Chincarini}, G., {Giommi}, P., {et~al.} 2004, \apj, 611, 1005

\bibitem[{{Gezari}(2021)}]{gezari2021}
{Gezari}, S. 2021, \araa, 59, 21

\bibitem[{{Gezari} {et~al.}(2017){Gezari}, {Cenko}, \& {Arcavi}}]{gezari2017}
{Gezari}, S., {Cenko}, S.~B., \& {Arcavi}, I. 2017, \apjl, 851, L47

\bibitem[{{Gomez} {et~al.}(2020){Gomez}, {Nicholl}, {Short}, {Margutti},
  {Alexander}, {Blanchard}, {Berger}, {Eftekhari}, {Schulze}, {Anderson},
  {Arcavi}, {Chornock}, {Cowperthwaite}, {Galbany}, {Herzog}, {Hiramatsu},
  {Hosseinzadeh}, {Laskar}, {M{\"u}ller Bravo}, {Patton}, \&
  {Terreran}}]{gomez2020}
{Gomez}, S., {Nicholl}, M., {Short}, P., {et~al.} 2020, \mnras, 497, 1925

\bibitem[{{Hernquist}(1990)}]{hernquist1990}
{Hernquist}, L. 1990, \apj, 356, 359

\bibitem[{{Hills}(1975)}]{hills1975}
{Hills}, J.~G. 1975, \nat, 254, 295

\bibitem[{{Holoien} {et~al.}(2016){Holoien}, {Kochanek}, {Prieto}, {Grupe},
  {Chen}, {Godoy-Rivera}, {Stanek}, {Shappee}, {Dong}, {Brown}, {Basu},
  {Beacom}, {Bersier}, {Brimacombe}, {Carlson}, {Falco}, {Johnston}, {Madore},
  {Pojmanski}, \& {Seibert}}]{holoien2016}
{Holoien}, T.~W.~S., {Kochanek}, C.~S., {Prieto}, J.~L., {et~al.} 2016, \mnras,
  463, 3813

\bibitem[{{Hu} \& {Loeb}(2021)}]{hu2021}
{Hu}, B.~X. \& {Loeb}, A. 2021, arXiv e-prints, arXiv:2105.14026

\bibitem[{{Hung} {et~al.}(2020){Hung}, {Foley}, {Ramirez-Ruiz}, {Dai},
  {Auchettl}, {Kilpatrick}, {Mockler}, {Brown}, {Coulter}, {Dimitriadis},
  {Holoien}, {Law-Smith}, {Piro}, {Rest}, {Rojas-Bravo}, \&
  {Siebert}}]{hung2020}
{Hung}, T., {Foley}, R.~J., {Ramirez-Ruiz}, E., {et~al.} 2020, \apj, 903, 31

\bibitem[{{Ivezi{\'c}} {et~al.}(2019){Ivezi{\'c}}, {Kahn}, {Tyson}, {Abel},
  {Acosta}, {Allsman}, {Alonso}, {AlSayyad}, {Anderson}, {Andrew}, {Angel},
  {Angeli}, {Ansari}, {Antilogus}, {Araujo}, {Armstrong}, {Arndt}, {Astier},
  {Aubourg}, {Auza}, {Axelrod}, {Bard}, {Barr}, {Barrau}, {Bartlett}, {Bauer},
  {Bauman}, {Baumont}, {Bechtol}, {Bechtol}, {Becker}, {Becla}, {Beldica},
  {Bellavia}, {Bianco}, {Biswas}, {Blanc}, {Blazek}, {Blandford}, {Bloom},
  {Bogart}, {Bond}, {Booth}, {Borgland}, {Borne}, {Bosch}, {Boutigny},
  {Brackett}, {Bradshaw}, {Brandt}, {Brown}, {Bullock}, {Burchat}, {Burke},
  {Cagnoli}, {Calabrese}, {Callahan}, {Callen}, {Carlin}, {Carlson},
  {Chandrasekharan}, {Charles-Emerson}, {Chesley}, {Cheu}, {Chiang}, {Chiang},
  {Chirino}, {Chow}, {Ciardi}, {Claver}, {Cohen-Tanugi}, {Cockrum}, {Coles},
  {Connolly}, {Cook}, {Cooray}, {Covey}, {Cribbs}, {Cui}, {Cutri}, {Daly},
  {Daniel}, {Daruich}, {Daubard}, {Daues}, {Dawson}, {Delgado}, {Dellapenna},
  {de Peyster}, {de Val-Borro}, {Digel}, {Doherty}, {Dubois},
  {Dubois-Felsmann}, {Durech}, {Economou}, {Eifler}, {Eracleous}, {Emmons},
  {Fausti Neto}, {Ferguson}, {Figueroa}, {Fisher-Levine}, {Focke}, {Foss},
  {Frank}, {Freemon}, {Gangler}, {Gawiser}, {Geary}, {Gee}, {Geha}, {Gessner},
  {Gibson}, {Gilmore}, {Glanzman}, {Glick}, {Goldina}, {Goldstein}, {Goodenow},
  {Graham}, {Gressler}, {Gris}, {Guy}, {Guyonnet}, {Haller}, {Harris},
  {Hascall}, {Haupt}, {Hernandez}, {Herrmann}, {Hileman}, {Hoblitt}, {Hodgson},
  {Hogan}, {Howard}, {Huang}, {Huffer}, {Ingraham}, {Innes}, {Jacoby}, {Jain},
  {Jammes}, {Jee}, {Jenness}, {Jernigan}, {Jevremovi{\'c}}, {Johns}, {Johnson},
  {Johnson}, {Jones}, {Juramy-Gilles}, {Juri{\'c}}, {Kalirai}, {Kallivayalil},
  {Kalmbach}, {Kantor}, {Karst}, {Kasliwal}, {Kelly}, {Kessler}, {Kinnison},
  {Kirkby}, {Knox}, {Kotov}, {Krabbendam}, {Krughoff}, {Kub{\'a}nek},
  {Kuczewski}, {Kulkarni}, {Ku}, {Kurita}, {Lage}, {Lambert}, {Lange},
  {Langton}, {Le Guillou}, {Levine}, {Liang}, {Lim}, {Lintott}, {Long},
  {Lopez}, {Lotz}, {Lupton}, {Lust}, {MacArthur}, {Mahabal}, {Mandelbaum},
  {Markiewicz}, {Marsh}, {Marshall}, {Marshall}, {May}, {McKercher}, {McQueen},
  {Meyers}, {Migliore}, {Miller}, {Mills}, {Miraval}, {Moeyens}, {Moolekamp},
  {Monet}, {Moniez}, {Monkewitz}, {Montgomery}, {Morrison}, {Mueller},
  {Muller}, {Mu{\~n}oz Arancibia}, {Neill}, {Newbry}, {Nief}, {Nomerotski},
  {Nordby}, {O'Connor}, {Oliver}, {Olivier}, {Olsen}, {O'Mullane}, {Ortiz},
  {Osier}, {Owen}, {Pain}, {Palecek}, {Parejko}, {Parsons}, {Pease},
  {Peterson}, {Peterson}, {Petravick}, {Libby Petrick}, {Petry},
  {Pierfederici}, {Pietrowicz}, {Pike}, {Pinto}, {Plante}, {Plate}, {Plutchak},
  {Price}, {Prouza}, {Radeka}, {Rajagopal}, {Rasmussen}, {Regnault}, {Reil},
  {Reiss}, {Reuter}, {Ridgway}, {Riot}, {Ritz}, {Robinson}, {Roby}, {Roodman},
  {Rosing}, {Roucelle}, {Rumore}, {Russo}, {Saha}, {Sassolas}, {Schalk},
  {Schellart}, {Schindler}, {Schmidt}, {Schneider}, {Schneider}, {Schoening},
  {Schumacher}, {Schwamb}, {Sebag}, {Selvy}, {Sembroski}, {Seppala}, {Serio},
  {Serrano}, {Shaw}, {Shipsey}, {Sick}, {Silvestri}, {Slater}, {Smith},
  {Smith}, {Sobhani}, {Soldahl}, {Storrie-Lombardi}, {Stover}, {Strauss},
  {Street}, {Stubbs}, {Sullivan}, {Sweeney}, {Swinbank}, {Szalay}, {Takacs},
  {Tether}, {Thaler}, {Thayer}, {Thomas}, {Thornton}, {Thukral}, {Tice},
  {Trilling}, {Turri}, {Van Berg}, {Vanden Berk}, {Vetter}, {Virieux},
  {Vucina}, {Wahl}, {Walkowicz}, {Walsh}, {Walter}, {Wang}, {Wang}, {Warner},
  {Wiecha}, {Willman}, {Winters}, {Wittman}, {Wolff}, {Wood-Vasey}, {Wu},
  {Xin}, {Yoachim}, \& {Zhan}}]{ivezic2019}
{Ivezi{\'c}}, {\v{Z}}., {Kahn}, S.~M., {Tyson}, J.~A., {et~al.} 2019, \apj,
  873, 111

\bibitem[{{Jose} {et~al.}(2014){Jose}, {Guo}, {Long}, {Herczeg}, {Dong},
  {Holoien}, {Prieto}, {Grupe}, {Shappee}, {Stanek}, {Kochanek}, {Davis},
  {Simonian}, {Basu}, {Beacom}, {Bersier}, {Brimacombe}, {Szczygiel}, \&
  {Pojmanski}}]{jose2014}
{Jose}, J., {Guo}, Z., {Long}, F., {et~al.} 2014, The Astronomer's Telegram,
  6777, 1

\bibitem[{{Kesden}(2012)}]{kesden2012}
{Kesden}, M. 2012, \prd, 85, 024037

\bibitem[{{Kochanek} {et~al.}(2017){Kochanek}, {Shappee}, {Stanek}, {Holoien},
  {Thompson}, {Prieto}, {Dong}, {Shields}, {Will}, {Britt}, {Perzanowski}, \&
  {Pojma{\'n}ski}}]{kochanek2017}
{Kochanek}, C.~S., {Shappee}, B.~J., {Stanek}, K.~Z., {et~al.} 2017, \pasp,
  129, 104502

\bibitem[{{Krolik} {et~al.}(2016){Krolik}, {Piran}, {Svirski}, \&
  {Cheng}}]{krolik2016}
{Krolik}, J., {Piran}, T., {Svirski}, G., \& {Cheng}, R.~M. 2016, \apj, 827,
  127

\bibitem[{{Lasota}(2023)}]{lasota2023}
{Lasota}, J.-P. 2023, arXiv e-prints, arXiv:2302.07925

\bibitem[{{Lidskii} \& {Ozernoi}(1979)}]{lidskii1979}
{Lidskii}, V.~V. \& {Ozernoi}, L.~M. 1979, Soviet Astronomy Letters, 5, 16

\bibitem[{{McConnell} \& {Ma}(2013)}]{mcconnell2013}
{McConnell}, N.~J. \& {Ma}, C.-P. 2013, \apj, 764, 184

\bibitem[{{Miniutti} {et~al.}(2023){Miniutti}, {Giustini}, {Arcodia}, {Saxton},
  {Read}, {Bianchi}, \& {Alexander}}]{miniutti2023}
{Miniutti}, G., {Giustini}, M., {Arcodia}, R., {et~al.} 2023, \aap, 670, A93

\bibitem[{{Mockler} {et~al.}(2019){Mockler}, {Guillochon}, \&
  {Ramirez-Ruiz}}]{mockler2019}
{Mockler}, B., {Guillochon}, J., \& {Ramirez-Ruiz}, E. 2019, \apj, 872, 151

\bibitem[{{Pasham} \& {van Velzen}(2018)}]{pasham2018}
{Pasham}, D.~R. \& {van Velzen}, S. 2018, \apj, 856, 1

\bibitem[{{Predehl} {et~al.}(2021){Predehl}, {Andritschke}, {Arefiev},
  {Babyshkin}, {Batanov}, {Becker}, {B{\"o}hringer}, {Bogomolov}, {Boller},
  {Borm}, {Bornemann}, {Br{\"a}uninger}, {Br{\"u}ggen}, {Brunner}, {Brusa},
  {Bulbul}, {Buntov}, {Burwitz}, {Burkert}, {Clerc}, {Churazov}, {Coutinho},
  {Dauser}, {Dennerl}, {Doroshenko}, {Eder}, {Emberger}, {Eraerds},
  {Finoguenov}, {Freyberg}, {Friedrich}, {Friedrich}, {F{\"u}rmetz},
  {Georgakakis}, {Gilfanov}, {Granato}, {Grossberger}, {Gueguen}, {Gureev},
  {Haberl}, {H{\"a}lker}, {Hartner}, {Hasinger}, {Huber}, {Ji}, {Kienlin},
  {Kink}, {Korotkov}, {Kreykenbohm}, {Lamer}, {Lomakin}, {Lapshov}, {Liu},
  {Maitra}, {Meidinger}, {Menz}, {Merloni}, {Mernik}, {Mican}, {Mohr},
  {M{\"u}ller}, {Nandra}, {Nazarov}, {Pacaud}, {Pavlinsky}, {Perinati},
  {Pfeffermann}, {Pietschner}, {Ramos-Ceja}, {Rau}, {Reiffers}, {Reiprich},
  {Robrade}, {Salvato}, {Sanders}, {Santangelo}, {Sasaki}, {Scheuerle},
  {Schmid}, {Schmitt}, {Schwope}, {Shirshakov}, {Steinmetz}, {Stewart},
  {Str{\"u}der}, {Sunyaev}, {Tenzer}, {Tiedemann}, {Tr{\"u}mper}, {Voron},
  {Weber}, {Wilms}, \& {Yaroshenko}}]{predehl2021}
{Predehl}, P., {Andritschke}, R., {Arefiev}, V., {et~al.} 2021, \aap, 647, A1

\bibitem[{Rees(1988)}]{rees1988}
Rees, M.~J. 1988, Nature, 333, 523

\bibitem[{{Roming} {et~al.}(2005){Roming}, {Kennedy}, {Mason}, {Nousek}, {Ahr},
  {Bingham}, {Broos}, {Carter}, {Hancock}, {Huckle}, {Hunsberger}, {Kawakami},
  {Killough}, {Koch}, {McLelland}, {Smith}, {Smith}, {Soto}, {Boyd},
  {Breeveld}, {Holland}, {Ivanushkina}, {Pryzby}, {Still}, \&
  {Stock}}]{roming2005}
{Roming}, P. W.~A., {Kennedy}, T.~E., {Mason}, K.~O., {et~al.} 2005, \ssr, 120,
  95

\bibitem[{Rubin \& Loeb(2011)}]{rubin2011}
Rubin, D. \& Loeb, A. 2011, Advances in Astronomy, 2011, 174105

\bibitem[{{Shakura} \& {Sunyaev}(1973)}]{shakura1973}
{Shakura}, N.~I. \& {Sunyaev}, R.~A. 1973, \aap, 24, 337

\bibitem[{{Shappee} {et~al.}(2014){Shappee}, {Prieto}, {Stanek}, {Kochanek},
  {Holoien}, {Jencson}, {Basu}, {Beacom}, {Szczygiel}, {Pojmanski},
  {Brimacombe}, {Dubberley}, {Elphick}, {Foale}, {Hawkins}, {Mullins},
  {Rosing}, {Ross}, \& {Walker}}]{shappee2014}
{Shappee}, B., {Prieto}, J., {Stanek}, K.~Z., {et~al.} 2014, in American
  Astronomical Society Meeting Abstracts, Vol. 223, American Astronomical
  Society Meeting Abstracts \#223, 236.03

\bibitem[{{Stone} {et~al.}(2019){Stone}, {Kesden}, {Cheng}, \& {van
  Velzen}}]{stone2019}
{Stone}, N.~C., {Kesden}, M., {Cheng}, R.~M., \& {van Velzen}, S. 2019, General
  Relativity and Gravitation, 51, 30

\bibitem[{{Tonry} {et~al.}(2018){Tonry}, {Denneau}, {Heinze}, {Stalder},
  {Smith}, {Smartt}, {Stubbs}, {Weiland}, \& {Rest}}]{tonry2018}
{Tonry}, J.~L., {Denneau}, L., {Heinze}, A.~N., {et~al.} 2018, \pasp, 130,
  064505

\bibitem[{{Tremaine} {et~al.}(1994){Tremaine}, {Richstone}, {Byun}, {Dressler},
  {Faber}, {Grillmair}, {Kormendy}, \& {Lauer}}]{tremaine1994}
{Tremaine}, S., {Richstone}, D.~O., {Byun}, Y.-I., {et~al.} 1994, \aj, 107, 634

\bibitem[{{van Velzen} {et~al.}(2016{\natexlab{a}}){van Velzen}, {Anderson},
  {Stone}, {Fraser}, {Wevers}, {Metzger}, {Jonker}, {van der Horst}, {Staley},
  {Mendez}, {Miller-Jones}, {Hodgkin}, {Campbell}, \& {Fender}}]{vanvelzen2016}
{van Velzen}, S., {Anderson}, G.~E., {Stone}, N.~C., {et~al.}
  2016{\natexlab{a}}, Science, 351, 62

\bibitem[{{van Velzen} {et~al.}(2016{\natexlab{b}}){van Velzen}, {Mendez},
  {Krolik}, \& {Gorjian}}]{vanvelzen2016b}
{van Velzen}, S., {Mendez}, A.~J., {Krolik}, J.~H., \& {Gorjian}, V.
  2016{\natexlab{b}}, \apj, 829, 19

\bibitem[{{Xiao} {et~al.}(2011){Xiao}, {Barth}, {Greene}, {Ho}, {Bentz},
  {Ludwig}, \& {Jiang}}]{xiao2011}
{Xiao}, T., {Barth}, A.~J., {Greene}, J.~E., {et~al.} 2011, \apj, 739, 28

\bibitem[{{Xie} \& {Yuan}(2012)}]{xie2012}
{Xie}, F.-G. \& {Yuan}, F. 2012, \mnras, 427, 1580

\bibitem[{{Yao} {et~al.}(2023){Yao}, {Ravi}, {Gezari}, {van Velzen}, {Lu},
  {Schulze}, {Somalwar}, {Kulkarni}, {Hammerstein}, {Nicholl}, {Graham},
  {Perley}, {Cenko}, {Stein}, {Ricarte}, {Chadayammuri}, {Quataert}, {Bellm},
  {Bloom}, {Dekany}, {Drake}, {Groom}, {Mahabal}, {Prince}, {Riddle},
  {Rusholme}, {Sharma}, {Sollerman}, \& {Yan}}]{yao2023}
{Yao}, Y., {Ravi}, V., {Gezari}, S., {et~al.} 2023, \apjl, 955, L6

\end{thebibliography}

\end{document}